
%
%
\def\today{\ifcase\month\or January\or February\or March\or April\or May\or
June\or July\or August\or September\or October\or November\or December\fi
\space\number\day, \number\year}
%
%
\newcount\notenumber

\def\note{\global\advance\notenumber by 1 \footnote{$^{\the\notenumber}$}}
%
%
\newif\ifsectionnumbering
\newcount\eqnumber
\def\cleareqnumber{\eqnumber=0}
\def\numbereq{\global\advance\eqnumber by 1
\ifsectionnumbering\eqno(\the\secnumber.\the\eqnumber)\else\eqno(\the\eqnumber)
\fi}
\def\eqalinno{{\global\advance\eqnumber by 1}
\ifsectionnumbering(\the\secnumber.\the\eqnumber)\else(\the\eqnumber)\fi}
\def\name#1{\ifsectionnumbering\xdef#1{\the\secnumber.\the\eqnumber}
\else\xdef#1{\the\eqnumber}\fi}

\sectionnumberingtrue
%
%
\newcount\refnumber

\immediate\openout1=refs.tex
\immediate\write1{\noexpand\frenchspacing}
\immediate\write1{\parskip=0pt}
\def\ref#1#2{\global\advance\refnumber by 1%
[\the\refnumber]\xdef#1{\the\refnumber}%
\immediate\write1{\noexpand\item{[#1]}#2}}
\def\tie{\noexpand~}

%
%
\font\twelvebf=cmbx10 scaled \magstep1
\newcount\secnumber

\def\newsection#1.{\ifsectionnumbering\cleareqnumber\else\fi%
	\global\advance\secnumber by 1%
	\bigbreak\bigskip\par%
	\line{\twelvebf \the\secnumber. #1.\hfil}\nobreak\medskip\par}
%
%
%
\def \sqr#1#2{{\vcenter{\vbox{\hrule height.#2pt
	\hbox{\vrule width.#2pt height#1pt \kern#1pt
		\vrule width.#2pt}
		\hrule height.#2pt}}}}

%
%
%
\newdimen\fullhsize
\def\fiddle{\fullhsize=6.5truein \hsize=3.2truein}
\def\fullline{\hbox to\fullhsize}
\def\mkhdline{\vbox to 0pt{\vskip-22.5pt
	\fullline{\vbox to8.5pt{}\the\headline}\vss}\nointerlineskip}
\def\mkftline{\baselineskip=24pt\fullline{\the\footline}}
\let\lr=L \newbox\leftcolumn
\def\twocolumns{\fiddle
	\output={\if L\lr \global\setbox\leftcolumn=\columnbox
		\global\let\lr=R \else \doubleformat \global\let\lr=L\fi
		\ifnum\outputpenalty>-20000 \else\dosupereject\fi}}
\def\doubleformat{\shipout\vbox{\mkhdline
		\fullline{\box\leftcolumn\hfil\columnbox}
		\mkftline} \advancepageno}
\def\columnbox{\leftline{\pagebody}}
\magnification=1200
\def\pr#1 {Phys. Rev. {\bf D#1\tie }}
\def\pe#1 {Phys. Rev. {\bf #1\tie}}
\def\pl#1 {Phys. Lett. {\bf #1B\tie }}
\def\prl#1 {Phys. Rev. Lett. {\bf #1\tie }}
\def\np#1 {Nucl. Phys. {\bf B#1\tie }}
\def\ap#1 {Ann. Phys. (NY) {\bf #1\tie }}
\def\cmp#1 {Commun. Math. Phys. {\bf #1\tie }}
\def\imp#1 {Int. Jour. Mod. Phys. {\bf A#1\tie }}
\def\mpl#1 {Mod. Phys. Lett. {\bf A#1\tie}}
\def\h{\textstyle{1\over 2}}
\def\tie{\noexpand~}

\parskip=15pt plus 4pt minus 3pt
\headline{\ifnum \pageno>1\it\hfil Number of Generations in
Free Fermionic String Models
	$\ldots$\else \hfil\fi}
\font\title=cmbx10 scaled\magstep1
\font\tit=cmti8 scaled\magstep1
\footline{\ifnum \pageno>1 \hfil \folio \hfil \else
\hfil\fi}
\raggedbottom
\rightline{\vbox{\hbox{CERN-TH.7518/94}\hbox{CTP-TAMU-64/94}\hbox{ACT-19/94}}}
\vfill
\centerline{\title NUMBER OF GENERATIONS IN FREE FERMIONIC}
\centerline{\title STRING MODELS}
\vfill
\centerline{\bf Ioannis Giannakis$^{(a),(b)}$, D.V.
Nanopoulos$^{(a),(b),(c)}$ and Kajia Yuan$^{(a),(b)}$}
\bigskip
\centerline{$^{(a)}${\tit Center for Theoretical Physics,
Texas A{\&}M University}}
\centerline{\tit College Station, TX 77843-4242, USA}
\medskip
\centerline{$^{(b)}${\tit Astroparticle Physics Group,
Houston Advanced Research Center (HARC)}}
\centerline{\tit The Mitchell Campus, The Woodlands, TX 77381, USA}
\medskip
\centerline{$^{(c)}${\tit CERN Theory Division, 1211 Geneva 23, Switzerland}}
\vfill
\centerline{\title Abstract}
\bigskip
{\narrower\narrower
In string theory there seems to be an intimate connection between
spacetime and world-sheet physics. Following this line
of thought we investigate the family problem in a particular class
of string solutions, namely the free fermionic string models. We
find that the number of generations $N_g$ is related to the index of the
supersymmetry generator of the underlying $N=2$ internal superconformal
field theory which is always present in any $N=1$ spacetime
supersymmetric string vacuum. We also derive a formula for the
index and thus for the number of generations which is sensitive
to the boundary condition assignments of the internal fermions
and to certain coefficients which determine the weight with
which each spin-structure of the model contributes to the
one-loop partition function. Finally we apply our formula to several
realistic string models in order to derive  $N_g$ and we verify
our results by constructing explicitly the massless spectrum
of these string models. \par}
\vfill\vfill\break

\newsection Introduction.

Recent experimental results indicate that the number of families
with light left-handed neutrinos is three. Although this result
suggests that additional families require neutrino components
with $m_{\nu_L}\geq 45$ GeV, the problem in understanding why
nature has chosen to fit the observed fermions into three generations
with identical quantum numbers remains. This is the so called
family problem and it is the modern version of Rabi's question
``Who ordered the muon?''.

Superstring theory is the only viable candidate for a consistent,
unified description of all interactions in nature. It is thus
interesting to explore the possible solution of the family
problem in this particular framework. Unfortunately although
great progress has been made towards understanding the structure
of string theory, it is fair to say that a full understanding is
still lacking. What we seem to understand in string theory is how
to construct
solutions of the theory: any two-dimensional conformal field
theory with the appropriate central charge represents a solution
of the theory. Of special importance is a class of conformal
field theories with $N=2$ world-sheet supersymmetry, since they
give rise to $N=1$ spacetime supersymmetry.
During the past several years a large number of solutions
of string theory has been constructed. Some of these solutions
can be viewed as compactifications of a ten-dimensional theory
(Calabi-Yau manifolds, orbifolds..) and as such have a geometric
interpretation while others are formulated directly in four
dimensions (free fermionic models, lattice constructions, $N=2$
superconformal models) with no apparent geometric interpretation.

In its present formulation string theory cannot adress the problem
of why nature has chosen to triplicate itself. At best we hope to
gain an insight into the problem by attempting to relate the
number of families of a particular string vacuum with some
characteristic of the underlying superconformal structure on the
world-sheet. In the Calabi-Yau
\ref{\witer}{P. Candelas, G. Horowitz, A. Strominger and E. Witten
\np258 (1985) 46.} and orbifold \ref{\wou}{L. Dixon, J. Harvey, C. Vafa
and E. Witten \np261 (1985) 678.} compactifications
the number of families is related to a topological quantity, the Euler
characteristic of the compactified internal space. Similar results
towards this direction have been obtained for L-G models in
\ref{\vaf}{C. Vafa \mpl12 (1989) 1169; K. Intriligator and
C. Vafa \np339 (1990) 95.}.
The free fermionic string models constitute a class of vacua of string
theory in which the extra degrees of freedom are realized in terms of
free world-sheet fermions. Subsequently these vacua lack a geometric
interpretation, although the information concerning the observable
four-dimensional physics remains encoded in the internal superconformal
field theory. As a result the family problem has remained
largely unexplored in this particular class of solutions despite
attempts to address it in \ref{\far}{A. Faraggi and D. V. Nanopoulos
\pr48 (1993) 3288.}.

It is the aim of this paper to address the family problem
in the free fermionic string models.
We will find that the number of families is given by the index of the
supersymmetry generator of the underlying $N=2$ internal superconformal
theory which is always present in any string vacuum with $N=1$
spacetime supersymmetry \ref{\cozu}{C. Hull and
E. Witten \pl160 (1985), 398; W. Boucher, D. Friedan and
A. Kent \pl172 (1986), 316; A. Sen \np278
(1986), 289; T. Banks, L. Dixon, D. Friedan and E. Martinec \np299 (1988),
613.}. In the next section we discuss the supersymmetric
index \ref{\bcx}{E. Witten \np202 (1982), 253.} of
an $N=2$ superconformal theory  \ref{\lerc}
{W. Lerche and N. Warner \pl205 (1988), 471.}. This index can
be expressed as a path integral over the degrees of freedom
of the theory. In section $3$ we will briefly review the free
fermionic string models in order to establish
the necessary notation. We will also derive an expression
for the index of the internal
$N=2$ superconformal field theory in terms of the contributions
of the internal fermions of the model to the partition function,
and provide arguments of why this formula gives us the
number of families of a particular string vacuum. The expression for
the index depends only on the boundary
condition assignments of the internal
fermions and a set of certain coefficients, which represent the weights
with which each sector of the string model contributes to the
one-loop string partition function.
Subsequently in section $4$ we will apply this formula to a
class of realistic free fermionic
string models with $N=1$ spacetime
supersymmetry and derive the number of
generations $N_g$ for these models. We will verify our results by
calculating explicitly the massless spectrum of these
models. Finally section $5$ will summarize our results.

\newsection The index of $N=2$ superconformal algebra.

An $N=2$ superconformal algebra is an extension of the Virasoro algebra.
In addition to the stress-energy tensor $T(z)$, its generators include
two spin-$3\over 2$ fields (the supercurrents) $T^{\pm}_{\rm F}(z)$,
and a spin-$1$ field $J(z)$, the generator of a $U(1)$ Kac-Moody
algebra. These fields can be expanded in terms of Laurent
modes as follows: $T(z)=\sum z^{-2-n}L_n$,
$T^{\pm}_{\rm F}(z)=\sum z^{-{3\over 2}-n\mp a}G^{\pm}_{n\pm a}$
and $J(z)=\sum z^{-1-n}J_n$, where $n\in {\bf Z}$, and the
parameter $a\in [0, 1)$ labels the boundary conditions of the
fermionic operators $T^{\pm}_{\rm F}$. In particular for the
Ramond (R) sector $a=0$, and for the Neveu-Schwarz (NS) sector
$a={1\over 2}$. The $N=2$ superconformal algebra can be expressed
either in terms of the operator product expansions of these fields
or equivalently in terms of the (anti)-commutation relations of their
modes. In terms of the modes the algebra is given
$$
\eqalign{
&\lbrack L_m, L_n \rbrack
=(m-n)L_{m+n}+{c\over 12}(m^3-m){\delta}_{m+n},
\quad \lbrack J_m, J_n \rbrack={c\over 3}m{\delta}_{m+n},
\cr
&\lbrace G^{+}_{m+a}, G^{-}_{n-a} \rbrace=L_{m+n}
+{1\over 2}(m-n+2a)J_{m+n}+{c\over 6}
[(m+a)^2-{1\over 4}]{\delta}_{m+n}, \cr
&\lbrack L_m, G^{\pm}_{n\pm a} \rbrack=
({m\over 2}-n\mp a)G^{\pm}_{m+n\pm a}, \quad
\lbrack L_m, J_n \rbrack=-nJ_{m+n}, \quad
\lbrack J_m, G^{\pm}_{n\pm a} \rbrack=\pm G^{\pm}_{m+n\pm a} \cr}
\numbereq\name{\eqrcop}
$$
Here $c$ is the central charge of the algebra.
For any given $a\in [0, 1)$, the complete representation
space of the corresponding $N=2$ superconformal algebra
can be obtained by acting
successively with $L_{-n}, J_{-n}, G_{-r}$ ($n, r > 0$) on the
highest weight states $|\phi\rangle$, states which
are annihilated by $L_n$,
$G_r$, $J_n$ for any $n, r > 0$ and are
eigenstates of the Cartan generators
$L_0$ and $J_0$ with eigenvalues $h$ and $q$ respectively:
$L_0|\phi\rangle =h|\phi\rangle$, $J_0|\phi\rangle=q|\phi\rangle$.

The fact that an $N=2$ superconformal algebra outlined above entails the
notion of an index can be most easily seen by considering the
subalgebra generated by $L_0$, $J_0$ and
$G^{\pm}_0$ ($G^{-}_0=G^{+\dagger}_0$) in the Ramond sector [\lerc],
where $h\geq {c\over 24}$ due to
unitarity constraints.
There are two types of states in the Hilbert space: (a)
massive states ($h> {c\over 24}$) which come in pairs
({\it i.e.} $|h,q\rangle$ and $G^{+}_0|h,q\rangle$)
with the same mass, since $\lbrack L_0, G^{\pm}_0 \rbrack=0$
and opposite eigenvalues of $(-1)^{J_0}$ since
$\lbrace (-1)^{J_0}, G^{\pm}_0 \rbrace=0$; and (b) massless states
($h={c\over 24}$) which are isolated. The latter are annihilated by
$G^{\pm}_0$: they are the supersymmetric ground states of the theory.
This situation is akin to the one which led Witten to introduce the
notion of the supersymmetric index in order to probe issues of
supersymmetry breaking[\bcx]. Then $G^{+}_0$ corresponds to the
``supersymmetry charge'' $Q$,
$L_0$ corresponds to the ``Hamiltonian'' $Q^2$,
and finally the operator $(-1)^{J_0}$ plays the role of
the ``chirality operator'' $(-1)^F$. Thus, for an $N=2$
superconformal algebra, one can define [\lerc] the index
of $G^{+}_0$ as ${\rm ind}\,(G^{+}_0)={\rm Tr}\,(-1)^{J_0}$.\note{For
any value $a\ne 0$, the index of the corresponding $N=2$
superconformal algebra can be defined via ``spectral flow''
\ref{\ss}{A. Schwimmer and N. Seiberg, \pl184 (1987) 191.}.}

There is one caveat in the above definition which is related with
the eigenvalues of $J_0$.
Strictly speaking, it is only when $(-1)^{J_0}=\pm 1$ that one can
meaningfully define $(-1)^{J_0}$ as a certain ``chirality operator''.
Under such condition, the representation space of the Ramond
sector decomposes into two subspaces $V_{+} \oplus V_{-}$ characterized
by the action of $(-1)^{J_0}$, with $(-1)^{J_0}V_{\pm}={\pm}V_{\pm}$.
As a consequence
the massless states split into those in
${\rm ker}\,(G^{+}_0)\subset V_{+}$,
which are annihilated by $G^{+}_0$ and those in
${\rm ker}\,(G^{-}_0)\subset V_{-}$, which are annihilated by $G^{-}_0$,
and ${\rm Tr}\,(-1)^{J_0}$ coincides with
${\rm dim\ ker}\,(G^{+}_0)-{\rm dim\ ker}\,(G^{-}_0)$.
The generic structure of the $N=2$ superconformal algebra alone
does not guarantee the eigenvalues of $J_0$ to be integers, so in
general ${\rm Tr}\,(-1)^{J_0}$ may be ill-defined as an index.
Fortunately, this caveat does not seem to present a real obstacle
for us. For heterotic string vacua, the presence of $N=1$ space time
supersymmetry requires that the local $(1, 0)$\note{In the convention
we use, the left-moving sector is supersymmetric,
whereas the right-moving sector is not.} superconformal symmetry
should be promoted to a global $(2, 0)$. If one considers
the total left-moving
$N=2$ superconformal algebra, which
contains both the space-time and internal degrees of freedom ($c=15$),
then the $U(1)$ charges (eigenvalues of $J_0$) of all physical
states are guaranteed to be integers. For reasons that will become
clear below, one often is interested in the $(2, 0)$ superconformal
theories ($c=9$) that describe only the internal degrees of freedom of
the classical vacua of the heterotic string. In that
case, the eigenvalues of $J_0$ are half
integers in the Ramond sector.
Neverthless, it is possible
to re-define the corresponding index ${\rm ind}\,(G^{+}_0)$ in
terms of the trace of an operator $(-1)^{J'_0}$ which
preserves the algebraic properties of $(-1)^{J_0}$ and has integer
eigenvalues. We will discuss an explicit example in the next section
when we consider the free fermionic string models. In the rest of this
section, we will write formally the index  operator
as the ${\rm Tr}(-1)^{J_0}$.
For applications in string theory, vacua with left-moving
$(2, 0)$ supersymmetry, the trace needs to be properly regularized
and the appropriate object to consider is [\lerc]
$$
{\rm ind}\,(G^{+}_0)={\rm Tr}\,[(-1)^{J_0}q^{(G^{+}_0+G^{-}_0)^2}
{\bar q}^{{\bar L}_0-{c\over 24}}]
={\rm Tr}\,[(-1)^{J_0}q^{L_0-{c\over 24}}
{\bar q}^{{\bar L}_0-{c\over 24}}],
\numbereq\name{\index2}
$$
where the trace runs over only the Ramond sector, $q=e^{2\pi i\tau}$
and ${\bar L}_0$ is the right-moving Hamiltonian.

The above discussion indicates that, for any heterotic string model
with $N=1$ spacetime supersymmetry, one can always define an index for
the supersymmetry generator of the underlying left-moving $N=2$
superconformal
algebra (\index2). We would like to
emphasize that this is true for the full class of heterotic string
vacua with $N=1$ space-time supersymmetry, which
includes both the familiar $(2, 2)$ models and the
more generic $(2, 0)$ ones. Furthermore,
in the current approach, the question of whether or
not the internal sectors of these string models admit
some geometric interpretation becomes secondary.

We now turn to the family problem in these string models.
Let us first recall that, although one speaks about light chiral
fermions, it is the space-time chirality
associated with the four-dimensional
Dirac operator that is really referred to, in string theory, the
consistency of the theory does establish a correlation between the
space-time chirality and
the internal chirality. In fact, the number of light chiral fermion
generations is given by the index of the internal Dirac operator
${\rm ind}\,({\not\!\!D})^{\rm int}={\rm Tr}\,{\Gamma}^{\rm int}$.
If the internal sector of the string model can be viewed as a compact
six-dimensional manifold $K$, (Calabi-Yau compactifications
with standard embedding and $(2, 2)$ symmetric
orbifold compactifications), one can further establish
a relation between the index of the internal
Dirac operator and the Euler characteristic $\chi (K)$ of $K$.
For more general string vacua whose internal sectors lack
any geometric interpretation, the concept of the Euler
characteristic becomes not directly
applicable. However, going back to the root of the problem, we see
that the crucial element here is the index of the
internal Dirac operator,
rather than the Euler characteristic. But how can one have a
handle on the former in general? Amazingly, the answer
to this question is provided
by the index of the $N=2$ superconformal algebra
discussed above. In fact,
it was shown in Ref.~[\lerc] that, {\it in the presence
of $N=1$ space-time supersymmetry}, the
index ${\rm Tr}\,(-1)^{J_0}$ is identical to the index
of the Dirac operator ${\rm Tr}\,\Gamma $. More specifically
the index ${\rm Tr}(-1)^{J_0}$ provides the number of
massless states of a given
eigenvalue of $(-1)^{J_0}$ minus the number of massless states of the
opposite eigenvalue. In a particular string vacuum $(-1)^{J_0}$
measures the internal chirality of a particle state. Consistency
of the theory imposes a correlation between $(-1)^{J_0}$ and the
chirality of a particle state as it is measured by an experimentalist
in our world. As a result the only contribution to the index arises
from chiral massless states (leptons, quarks..) while massless
vector-like particles (Higgs, gravitinos, gauginos..) do not
contribute.
Since for string models with
$N=1$ space-time supersymmetry their internal sectors realize
an $N=2$ superconformal algebra ($(2, 0)$ world-sheet supersymmetry)
with $c=9$, the natural object to consider therefore is the index
(\index2) for the internal sectors.

\newsection Index formula in free fermionic string models.

In the previous section, we have associated the number of families
in any $N=1$ spacetime supersymmetric four-dimensional heterotic
string model with the index (\index2) of the underlying
{\it internal} left-moving $N=2$ superconformal
algebra. We now focus on a particular class of such four-dimensional
heterotic string models, namely the free fermionic string models
\ref{\KLT}{H. Kawai, D.C. Lewellen and H. Tye, \prl57 (1986) 1832;
\np288 (1987) 1.}\ref{\AB}{I. Antoniadis, C. Bachas and C. Kounnas,
\np289 (1987) 87; I. Antoniadis and C. Bachas, \np298 (1988) 586.}\ref
{\BD}{R. Bluhm, L. Dolan and P. Goddard, \np309 (1988) 330.}.
In the free fermionic constructions the internal degrees of freedom
are realized in terms of free world-sheet fermions.
In the light-cone gauge, the two transverse space-time coordinates
$X^\mu$ and their left-moving superpartners ${\psi}^\mu$ are
supplemented by real or complex left and right-moving free
world-sheet fermions.
These fermionic degrees of freedom furnish a representation
of an internal left-moving
$N=1$ superconformal theory with $c_L=9$ and a right-moving
conformal theory with $c_R=22$. For instance, in
a particularly interesting
class of models, the left-moving internal sector can
be specified in terms
of the $18$ real fermions being arranged into six sets
$(\chi^I, y^I, \omega^I)$
($I=1,\cdots,6$), each transforming in the adjoint
representation of
$SU(2)$, such that the local world-sheet supersymmetry is
non-linearly realized among them. The supercurrent is
then given by the following expression
\ref{\ellh}
{I. Antoniadis, C. Bachas, C. Kounnas and P. Windey \pl171 (1986), 51.}:
$$
T_{\rm F}(z)={\psi^{\mu}}{\partial}{X_{\mu}}+i{\sum_I}{\chi^I}y^I{\omega^I}.
\numbereq\name{\eqgort}
$$
As for the right-moving internal sector, in general one can use $n_r$
real fermions and $n_c$ complex fermions, as long as $n_r+2n_c=44$.

The essential ingredient in the free fermionic formulation is the
spin-structures of all the fermions,
namely the possible boundary condition assignments for the fermions
as they are parallel transported around the two non-contractible
loops ($\sigma, \tau$ directions) of a torus. It is convenient
to use a vector, which we refer to as a spin-structure vector, to
represent a set of boundary conditions for all the world-sheet fermions.
We follow the notation of the second paper in Ref.~[\AB].
For instance, an entry 1(0) in spin-structure vectors means that the
corresponding fermion has periodic (anti-periodic) boundary condition.
In this construction, a particular string model is completely
specified by:
(a) a basis ${\cal B}=\lbrace b_1, b_2, \ldots b_n \rbrace $ of
some vectors which generates
all consistent spin-structures of the model.
These vectors altogether form a finite additive group $\Xi$ of
dimension $M=\prod^n_{i=1}\,N_{b_i}$, where $N_{b_i}$ is the order
of the basis vector $b_i$ and (b) a set of spin-structure coefficients
$C{{b_i}\atopwithdelims[] {b_j}}$ for a pair of basis vectors
$b_i, b_j\in {\cal B}$, which determines the relative weights with
which all possible spin-structures contribute to the one-loop string
partition function. Consistency conditions
of string theory, such as modular
invariance {\it etc.}, can be cast into a set of simple rules for the
basis vectors $b_i\in {\cal B}$ and the coefficients
$C{{b_i}\atopwithdelims[] {b_j}}$ [\KLT][\AB].

The total Hilbert space of the string model consists of
$M$ sectors, each corresponding to an
element of $\Xi$, {\it i.e.},  for every $\alpha\in \Xi$ there
is a $\alpha$-sector. The set of all sectors can be divided into two
classes $\Xi^{+}$ and $\Xi^{-}$, each containing
$M/2$ sectors, depending on the
value of $\delta_\alpha=e^{i\pi\alpha(\psi^\mu)}=\pm 1$.
Since the bosonic or fermionic nature of the particle state is solely
determined by the boundary conditions of the world-sheet fermions
with spacetime index $\psi^\mu$, if $\alpha\in\Xi^{+}$, the
$\alpha$-sector provides space-time bosons, while if $\alpha\in\Xi^{-}$,
it provides space-time fermions.
In a given $\alpha$-sector, the spin-structures have the same assignment
in the $\sigma$-direction (around the string) determined
by the vector $\alpha$, while their assignments along the
$\tau$-direction can still be allowed to vary according
to any vector $\beta\in \Xi$. For every $\alpha$-sector, the summation
over contributions to one-loop string partition function from all
such $\beta$ in $\tau$-direction realizes a GSO-type projection, which
is equivalent to a set of separate GSO-projections, one for each basis
vector. Finally, the GSO-projections for a specific $\alpha$-sector
determines the physical states in that sector.

In order to consturct a fermionic string model with $N=1$ space-time
supersymmetry, it is important to have sectors that could give
rise to massless gravitinos. This can be achieved by including one
such vector ($S$-vectors)~\ref{\REISS}
{H. Dreiner, J.L. Lopez, D.V. Nanopoulos and D. Reiss, \np320 (1989) 401.}
in our basis. The simplest choice is given
by
$$
S=(1\ 100\ 100\ 100\ 100\ 100\ 100 : {\bf 0}_R). \numbereq\name{\svec}
$$
In our notation, the first entry is reserved for the two transverse
$\psi^\mu$ treated as a single complex fermion, the following 18
entries correspond to the left-moving internal fermions
$(\chi^I, y^I, \omega^I)$. This is the only $S$-vector which permits the
realization of the left-moving internal sector entirely in terms
of real fermions,
all other $S$-vectors given in Ref.~[\REISS] require the use of complex
fermions. This particular choice for the $S$-vector only
guarantees that the
model will have $N=4$ space-time supersymmetry. Subsequently one
needs to introduce other basis vectors
in order to reduce space-time supersymmetry
from $N=4$ to $N=1$. We would like to emphasize that the choice
of coefficients $C{{S}\atopwithdelims[] {b_i}}$ for any
$b_i\in {\cal B}$ is also very crucial, because it is possible that
even if the model has the correct basis vectors
in order to generate $N=1$
space-time supersymmetry, some otherwise consistent choice of
certain $C{{S}\atopwithdelims[] {b_i}}$ could result in the elimination
of the gravitinos from the spectrum through the GSO-projections.
There is only a limited number of
choices for the $C{{S}\atopwithdelims[] {b_i}}$ that one can make. In
fact, given the above $S$-vector, the sufficient condition for
$N=1$ supersymmetry is \ref{\kajia}{K. Yuan, Ph.D. Thesis, Texas A\&M
University (1991).}
$$
C{{S}\atopwithdelims[] {b_i}}=-\delta_{b_i}. \numbereq\name{\choice}
$$
This condition in the free fermionic formulation is analogous to
the condition for the vanishing of the first Chern class in Calabi-Yau
or orbifold compactifications.

For free fermionic models with $N=1$ space-time supersymmetry,
the left-moving local $(1, 0)$ superconformal symmetry can
be extended to a global $(2, 0)$ one. In general,
the precise realization of this $(2, 0)$ structure in the fermionic
models is contigent on some details of the models, the key issue being
how to construct the $U(1)$ current $J(z)$ in terms of the world-sheet
fermions. With the $S$-vector given as (\svec), the $U(1)$ current
$J(z)$ of the {\it internal} left-moving $N=2$ superconformal algebra
can be constructed in terms of the six periodic fermions $\chi^I$ in
the $S$-vector, as follows \ref{\nanop}{S. Kalara, J.L. Lopez and
D.V. Nanopoulos, \np353 (1991) 650.}
$$
J(z)=i\partial_z(S_{12}+S_{34}+S_{56}), \numbereq\name{\uone}
$$
where $S_{12}$ is the bosonization of the complex fermion $\chi^{12}$
defined as $e^{iS_{12}}=\chi^{12}={1\over {\sqrt 2}}(\chi^1+i\chi^2)$,
and similarly for $S_{34}$ and $S_{56}$. The internal part of the
supercurrent $T_{\rm F}$ (see Eq.~(\eqgort)) indeed splits into
two pieces $T^{\pm}_{\rm F}$ with eigenvalues $\pm 1$ under the
action of $J(z)$. Since in the Ramond sector of this internal
$N=2$ superconformal algebra $T^{\pm}_{\rm F}$ are
periodic,\note{We remind the cautious readers that the usual analysis
of fermionic formulation is carried out on the cylinder,
whereas the presentation of the previous
section is given on the complex plane.}
from Eq.~(\eqgort) we see that in this sector $\psi^\mu$ should
also be periodic. Therefore, in studying the index (\index2) of the
internal $N=2$ superconformal algebra for this class of fermionic models,
we only need to include the $\alpha$-sectors with $\delta_\alpha=-1$,
namely, those $\alpha\in\Xi^{-}$.

However, for these sectors the eigenvalues of $J_0$ are
half-integers instead of integers [\nanop]. To see why this is
the case, we first note that the eigenvalues of $J_0$ are
the sum of the so-called fermionic charges [\KLT] of the three
complex fermions $\chi^{12}, \chi^{34}$ and $\chi^{56}$.
We recall that the fermionic charge $Q_\alpha(f)$
of a complex fermion $f$ in a particular $\alpha$-sector is
given by [\KLT]\ref{\fli1}{I. Antoniadis, J. Ellis, J. Hagelin and
D.V. Nanopoulos, \pl205 (1988) 459.}
$$
Q_\alpha(f)=F_\alpha(f)+{1\over 2}\alpha(f), \numbereq\name{\charge}
$$
where $F_\alpha(f)$ is the fermion number
and the second term represents the vacuum
charge arising from filling the negative energy states. Thus, it
is the vacuum charges involved in the definition of $J(z)$ which
cause the problem here. Clearly then we can regulate
$J_0$ by removing the vacuum charges, the proper operator $J'_0$ that
we alluded to in the previous section is then simply
the sum of the fermion
numbers of $\chi^{12}, \chi^{34}$ and $\chi^{56}$.

We now proceed with the computation of the index ${\rm Tr}\,(-1)^{J'_0}$
for free fermionic string models.
To this end let us first recall that the
index ${\rm Tr}\,(-1)^F$ of a general $1+1$ dimensional supersymmetric
non-linear $\sigma$-model [\bcx] can be computed as a path
integral \ref{\breq}{S. Cecotti and L. Girardello, \pl110 (1982) 253.}
$$
{\rm Tr}\,(-1)^F={\int}dX^I(\sigma, \tau)\,d{\psi}^I(\sigma, \tau)\,e^{-A},
\numbereq\name{\eqroiuv}
$$
where $X^I(\sigma, \tau)$ represent the bosonic degrees of freedom and
$\psi^I(\sigma, \tau)$ their superpartners. The integration is over
periodic fermionic degrees of freedom in both $\sigma$ and $\tau$
directions due to the insertion of $(-1)^F$, and $A$ is
the action of the supersymmetric non-linear $\sigma$-model.
We note that in
(\eqroiuv), $X^I$ and $\psi^I$ include both left-moving and
right-moving degrees of freedom. Whenever it is possible to
interpret the bosonic degrees of freedom $X^I$ of the
non-linear sigma model as coordinates of a compact
six-dimensional manifold $K$, as it is the case with Calabi-Yau and
orbifold compactifications with standard embedding, equation (\eqroiuv)
gives the Euler characteristic of $K$.
However, ever since the advent of heterotic string theory,
it became evident that string theory {\it per se} does not
demand a symmetrical treatment between the left and right degrees
of freedom. This is true for the asymmetric
orbifolds \ref{\asb}{K. Narain, M. Sarmadi and C. Vafa, \np288 (1987) 551.}
as well as for the free fermionic string models. Therefore we are
led to a generalization of (\eqroiuv)
given by
$$
{\rm Tr}\,(-1)^{J'_0}={\int}dX^I_L\,dX^I_R\,d{\psi}^I_L\,d{\psi}^I_R\,e^{-A}.
\numbereq\name{\path2}
$$
We would like to emphasize that (\path2) indeed generalizes (\eqroiuv).
The point is that, since in (\path2) the operator $J'_0$ is defined
interms of $\psi^I_L$, only the left-moving world-sheet supersymmetry
is needed. In other words, it is not necessary to have right-moving
supersymmetry among $X^I_R$ and $\psi^I_R$ for (\path2) to make sense.
In short, (\path2) is valid for any $(2, 0)$ models, while (\eqroiuv)
is valid only for $(2, 2)$ models.

For free fermionic models, the fields $X^I_L$ and $\psi^I_L$ in (\path2)
obviously correspond to the 18 left-moving fermions
$(\chi^I, y^I, \omega^I)$. In fact, given the form of the $S$-vector,
it is easy to see that each pair $(y^I, \omega^I)$ can be viewed as
the fermionization of a bosonic field $X^I_L$, and $\chi^I$ are simply
the corresponding $\psi^I_L$. What appears to be uncertain is the
representation of the right-moving degrees of freedom
$X^I_R$ and $\psi^I_R$ in terms of some right-moving world-sheet
fermions. If $m_r$ real fermions and $m_c$ complex fermions
are included, one must have $m_r+2m_c=18$ to match the right-moving
degrees of freedom. Moreover, for a fermionic
model with a $(2, 2)$ superconformal structure, the right-moving
fermions in (\path2) obviously should be those which realize the
global right-moving $N=2$ superconformal algebra, because in that
case (\path2) is identical to (\eqroiuv).
This observation enables one to determine these internal right-moving
fermions in some $(2, 0)$ models as well. Suppose one is given
a fermionic model with a basis ${\cal B}$, which has $N=1$ space-time
supersymmetry but only $(2, 0)$ world-sheet supersymmetry. One then
can investigate all the normal subsets of ${\cal B}$ that
would also give rise to models with $N=1$ space-time supersymmetry. If
one can find a sub-model which actually admits a $(2, 2)$ structure,
then the right-moving fermions that are responsible for the presence of
the additional $(0, 2)$ structure in this sub-model are the
ones that should be
included in (\path2) for the index of the original $(2, 0)$ model.
This procedure may sound fairly long-winded at this
point but, in reality,
it is not. We will provide some examples in the next section
in order to demonstrate
how this is done in practice. Of course, we are aware of the fact that
maybe not all of the $(2, 0)$ fermionic models are
bound to have a $(2, 2)$
``core'' in the sense we have just described. In
that case, the problem of
deciding which right-moving fermions should be included in (\path2)
remains.

Having specified all the fields in (\path2), we can proceed to
compute the
path integral. The calclulation is very similar to that of
the one-loop fermionic
partition function [\KLT][\AB]. As in the case
of orbifolds, in the free
fermionic models one is bounded to integrate over field configurations
with twisted boundary conditions, namely over
different spin-structures for the appropriate fermions.
In other words, the index ${\rm Tr}\,(-1)^{J'_0}$ is in this case
a {\it character-valued index} ${\rm Tr}\,g(-1)^{J'_0}$,
where $g$ twists the boundary conditions of the internal world-sheet
fermions in both $\sigma$ and $\tau$ directions. Since only massless
states contribute to the index, the total contribution from
different $\sigma$-twists manifests itself as a sum
over all {\it massless}
$\alpha$-sectors with $\alpha\in\Xi^{-}$. For each
of such $\alpha$-sector,
similar to the orbifold case, the contributions
from the $\tau$-twists form a
projection onto invariant states. The
insertion of $(-1)^{J'_0}$ further
implies that the only $\tau$-twists which give non-zero
contribution are those given by the
vectors $\beta\in\Xi^{-}$. There are $M/2$ such $\beta$ vectors, so
the projection amounts to a summation over them divided by $M/2$.
In addition, we expect the index ${\rm Tr}\,(-1)^{J'_0}$
being a topological quantity, to remain invariant under
modular transformations.
Putting all these pieces together, we derive a formula for
the index of the underlying $N=2$ superconformal theory in
the free fermionic string models given by
$$
{\rm Tr}\,(-1)^{J'_0}={2\over {\prod_i}N_{b_i}}
{\sum_{\alpha\in \Xi^{-}}}'\,
{\sum_{\beta\in \Xi^{-}}}\,C{{\alpha} \atopwithdelims[] {\beta}}
Z^{\rm int}{{\alpha} \atopwithdelims[] {\beta}}.
\numbereq\name{\eqboria}
$$
Here the prime on $\alpha$-summation sign indicates that
only massless Ramond sectors contribute. The spin-structure
coefficients $C{\alpha \atopwithdelims[] \beta}$ in (\eqboria)
are the same as those appearing in the one-loop
string partition function,
which can be completely determined by the basic set of
$C{{b_i} \atopwithdelims[] {b_j}}$. Finally,
$Z^{\rm int}{{\alpha} \atopwithdelims[] {\beta}}$ represents
the product of all determinants of the world-sheet Dirac operators
for the internal fermions, with $\sigma$- and $\tau$-boundary
conditions specified by $\alpha$ and $\beta$, respectively.
Using our phase convention, the determinant of a single
complex left-moving
fermion $f$ with $\alpha(f)=a$ and $\beta(f)=b$ is given in
terms of the $\Theta$ function as
$$
Z{a \atopwithdelims() b}={e^{i\pi (ab)/2}\over \eta(\tau)}\,
\Theta{{a/2} \atopwithdelims[] -{b/2}}(0|\tau),
\numbereq\name{\sita}
$$
where $\eta(\tau)$ is the Dedekind function. The determinant of a
single complex right-moving fermion is obtained by taking the complex
conjugate of (\sita). For real fermions, one needs to take the square
root of these expressions. The right hand side of the index formula
(\eqboria) can be either positive or negative and as we have argued
in the previous section it provides a measure of the space-time chiral
asymmetry in the free fermionic string models.
Its absolute value then provides the number of generations $N_{\rm gen}$
in such models.

\newsection Some illustrative examples.

The purpose of this
section is two-fold. First, we would like to apply the index
formula (\eqboria) presented in the previous  section
in particular string models and demonstrate explicitly how to choose
the internal right-moving
fermions and second we would like to verify the results obtained in
this manner by comparing them with the results which
arise by direct derivation of the massless spectrum of these models.

Let us first start with a simple basis
${\cal B}_1=\lbrace {\bf 1}, S, b_1, b_2, b_3 \rbrace$, where in
vector ${\bf 1}$ all world-sheet fermions are periodic, vector
$S$ is given by (\svec) and
$$
\eqalign{
b_1&=(1\ 100\ 100\ 010\ 010\ 010\ 010 : \overbrace{001111}^{{\bar y}^I}
\ \!\overbrace{000000}^{{\bar \omega}^I}\ \!\!\overbrace{100}^{{\bar \eta}^k}
\ \!\!\overbrace{11111}^{{\bar \psi}^a}\ {\bf 0}_8),\cr
b_2&=(1\ 010\ 010\ 100\ 100\ 001\ 001 : 110000\ 000011\ 010\ 11111\
{\bf 0}_8),\cr
b_3&=(1\ 001\ 001\ 001\ 001\ 100\ 100 : 000000\ 111100\ 001\ 11111\
{\bf 0}_8).\cr} \numbereq\name{\eqfopm}
$$
This set of basis vectors was first introduced in the construction of
the flipped $SU(5)$ model in Ref.~[\fli1] and have been adopted in many
other free fermionic models. In writing these spin-structure
vectors we have separated the right-moving fermions into 12 real
fermions ${\bar y}^I, {\bar \omega}^I$ $(I=1,\cdots,6)$ while the rest
are treated as complex fermions. In these vectors, the
boundary conditions
for each pair $({\bar y}^I, {\bar \omega}^I)$ are
identical to those for
the corresponding pair $(y^I, \omega^I)$. Similarly,
right-moving complex
fermions ${\bar \eta}^{1,2,3}$ have the same boundary conditions as
left-moving fermions $\chi^{12,34,56}$ in all vectors except for $S$.
This simple observation prompts us to include these right-moving
fermions ($m_r=12, m_c=3$) in the index formula (\eqboria). In fact,
it is easy to see that these fermions indeed realize
an internal right-moving
$N=2$ superconformal algebra, {\it i.e.}, an additional global $(0, 2)$
world-sheet supersymmetry. Upon bosonization the 12
real fermions $({\bar y}^I, {\bar \omega}^I)$ provide the right-moving
parts $X^I_R$ of the coordinates of the internal six-dimensional
manifold $K$ while the ${\bar \eta}^{1,2,3}$ correspond to the degrees
of freedom that realize the embedding of the spin connection into the
gauge group.

In any fermionic model generated by the basis ${\cal B}_1$
there are $M/2=2^5/2=16$ vectors in $\Xi^{-}$, since
$N_{b_i}=2$ for any $b_i\in {\cal B}_1$. These
provide the following 16 sectors:
$$
\eqalign{
&{\bf 1}\ (+\zeta),\ \ S\ (+\zeta), \cr
&b_1\ (+\zeta),\ \ b_2\ (+\zeta),\ \ b_3\ (+\zeta), \cr
&S+b_1+b_2\ (+\zeta),\ \ S+b_2+b_3\ (+\zeta),\ \ S+b_3+b_1\ (+\zeta).\cr}
\numbereq\name{\eqseta}
$$
Here for convenience we have introduced $\zeta\equiv {\bf 1}+b_1+b_2+b_3$.
The boundary conditions of the internal fermions in a particular vector
are not affected by adding $\zeta$, thus
$$
Z^{\rm int}{{\alpha} \atopwithdelims[] {\beta}}
=Z^{\rm int}{{\alpha+\zeta} \atopwithdelims[] {\beta}}
=Z^{\rm int}{{\alpha} \atopwithdelims[] {\beta+\zeta}}
=Z^{\rm int}{{\alpha+\zeta} \atopwithdelims[] {\beta+\zeta}}.
\numbereq\name{\easy}
$$
Among the 16 sectors in (\eqseta) which give rise to
space-time fermions,
only $(S,S+\zeta,b_1,b_2,b_3)$ will produce massless fermions, so
there are five terms in the $\alpha$-summation of the index
formula (\eqboria). First, for any vector $\beta$ taken from the set
(\eqseta), we have
$$
Z^{\rm int}{{S} \atopwithdelims[] {\beta}}=0.
\numbereq\name{\whys}
$$
This is because $Z^{\rm int}{{S} \atopwithdelims[] {\beta}}$
always includes at least one factor $Z{1 \atopwithdelims() 1}$
(one left-moving real fermion with periodic boundary conditions in
both $\sigma$ and $\tau$ directions) which vanishes according to
Eq.(\sita)
$$
Z{1 \atopwithdelims() 1}
=e^{i\pi /4}\sqrt{{\theta_1(\tau)\over \eta(\tau)}}=0.
\numbereq\name{\kigo}
$$
Therefore, the $S$-sector does not contribute
to the index. The same is also
true for the $(S+\zeta)$-sector because of (\easy). In fact,
we know that
these two sectors provide space-time gravitinos
and gauginos which are not
chiral. In bosonic language these two sectors correspond to the
untwisted sectors of the orbifold compactifications.
This is in agreement with the fact that chiral fermions in the
orbifold models arise from the twisted sectors.
We would like to point out that, except
for the condition (\choice) which guarantees the validity of the
$N=2$ index itself, this particular result for
the sectors $S$ and $(S+\zeta)$
is independent of the choice of the coefficients
$C{{b_i} \atopwithdelims[] {b_j}}$, therefore, it remains true
for all models generated by
the basis ${\cal B}_1$ with $N=1$ space-time
supersymmetry.

We next calculate the contributions from the sectors ($b_1, b_2, b_3$).
Several terms in the $\beta$-summation vanish  due to Eq. (\kigo).
For instance, making use of
(\easy), the contribution of the $b_1$-sector
is given by
$$
I_{b_1}={1\over 16}\biggl (\,C{{b_1} \atopwithdelims[] {b_2}}
+C{{b_1} \atopwithdelims[] {b_2+\zeta}}\,\biggr )\,
Z^{\rm int}{{b_1} \atopwithdelims[] {b_2}}+
{1\over 16}\biggl (\,C{{b_1} \atopwithdelims[] {b_3}}
+C{{b_1} \atopwithdelims[] {b_3+\zeta}}\,\biggr )\,
Z^{\rm int}{{b_1} \atopwithdelims[] {b_3}}.
\numbereq\name{\eqccb1}
$$
To compute this expression, we write the Dirac determinants
in terms of the $\theta$-functions according to the following relations
(see Eq.~(\sita))
$$
Z{1 \atopwithdelims() 0}=\sqrt{{\theta_2(\tau)\over \eta(\tau)}},\quad
Z{0 \atopwithdelims() 0}=\sqrt{{\theta_3(\tau)\over \eta(\tau)}},\quad
Z{0 \atopwithdelims() 1}=\sqrt{{\theta_4(\tau)\over \eta(\tau)}}
\numbereq\name{\eqfois}
$$
and then use the well-known Jacobi triplet product identity
$$
\theta_2(\tau)\theta_3(\tau)\theta_4(\tau)=2\eta^3(\tau)
\numbereq\name{\eqvopus}
$$
The result is then (note ${\bar \eta}^{1,2,3}$ are complex)
$$
Z^{\rm int}{{b_1} \atopwithdelims[] {b_2}}=
Z^{\rm int}{{b_1} \atopwithdelims[] {b_3}}=2^6.
\numbereq\name{\eqzzno}
$$
As for the spin-structure coefficients, using the rules given in
Ref.~[\AB], we immediately get
$$
C{{b_1} \atopwithdelims[] {b_2+\zeta}}=C{{b_1} \atopwithdelims[] {b_3}},\quad
C{{b_1} \atopwithdelims[] {b_3+\zeta}}=C{{b_1} \atopwithdelims[] {b_2}}
\numbereq\name{\eqrvuo}
$$
Therefore, Eq.~(\eqccb1) reduces to
$$
I_{b_1}=8\biggl (\,C{{b_1} \atopwithdelims[] {b_2}}
+C{{b_1} \atopwithdelims[] {b_3}}\,\biggr )
\numbereq\name{\eqcb1}
$$
Similarly, we have
$$
I_{b_2}=8\biggl (\,C{{b_2} \atopwithdelims[] {b_3}}
+C{{b_2} \atopwithdelims[] {b_1}}\,\biggr ),\quad
I_{b_3}=8\biggl (\,C{{b_3} \atopwithdelims[] {b_1}}
+C{{b_3} \atopwithdelims[] {b_2}}\,\biggr ),
\numbereq\name{\eqccb23}
$$
for the $b_2$- and $b_3$-sector, respectively.

So far, we have not yet specified the basic set of the coefficients
$C{{b_i} \atopwithdelims[] {b_j}}$, except for
$C{S \atopwithdelims[] {b_i}}$ which were chosen to satisfy (\choice).
After we fix $C{{\bf 1} \atopwithdelims[] {\bf 1}}=-1$, there are
still six coefficients that can be independently chosen to be $\pm 1$.
However, it turns out that the three coefficients
$C{{b_1} \atopwithdelims[] {\bf 1}}$, $C{{b_2} \atopwithdelims[] {\bf 1}}$
and $C{{b_3} \atopwithdelims[] {\bf 1}}$ have no effects on the index
of the $N=2$ supercoformal theory.
So we are left with eight possibilities given by the values of the
three coefficients $C{{b_1} \atopwithdelims[] {b_2}}$,
$C{{b_1} \atopwithdelims[] {b_3}}$ and $C{{b_2} \atopwithdelims[] {b_3}}$.
These will give rise to eight models, but as far as the
chiral asymmetry is concerned, there could be only two
cases: (a) all these three coefficients
are chosen to be the same; and (b) only two of these
coefficients are
the same, the third one differs by a sign.
As an example of the first case, in our first model we choose
$$
C{{b_1} \atopwithdelims[] {b_2}}=
C{{b_1} \atopwithdelims[] {b_3}}=
C{{b_2} \atopwithdelims[] {b_3}}=-1 \numbereq\name{\eqmma}
$$
{}From (\eqcb1) and (\eqccb23), in this model we get
$I_{b_1}=I_{b_2}=I_{b_3}=-16$, so the $N=2$ index is
$$
{\rm Tr}\,(-1)^{J'_0}=I_{b_1}+I_{b_2}+I_{b_3}=-48
\numbereq\name{\eqrcls}
$$
This indicates that there are 48 generations in this model, a result
that one could also derive by working out the massless spectrum of the
model explicitly.
In fact, in this model, each of the $b_1$, $b_2$ and $b_3$
produces two copies
of space-time fermions, each copy transforming under the gauge group
$SO(10)\times{SO(6)^3}\times E_8$ as follows
$$
\eqalign{
&b_1\,:\ ({\bf 16}, {\bf 4}, {\bf 1}, {\bf 1}, {\bf 1})+
({\bf 16}, {\bf {\bar 4}}, {\bf 1}, {\bf 1}, {\bf 1})\cr
&b_2\,:\ ({\bf 16}, {\bf 1}, {\bf 4}, {\bf 1}, {\bf 1})+
({\bf 16}, {\bf 1}, {\bf {\bar 4}}, {\bf 1}, {\bf 1})\cr
&b_3\,:\ ({\bf 16}, {\bf 1}, {\bf 1}, {\bf 4}, {\bf 1})+
({\bf 16}, {\bf 1}, {\bf 1}, {\bf {\bar 4}}, {\bf 1})\cr}
\numbereq\name{\eqcosa}
$$
Each of $b_1$, $b_2$ and $b_3$ provides us with $2(4+4)=16$
generations of chiral fermions which transform as ${\bf 16}$
of $SO(10)$, and as a result we have a model with 48 families in
total. It is interesting to note from the previous analysis that,
$I_{b_1}$, $I_{b_2}$ and $I_{b_3}$, {\it i.e.}, the contributions
from the $b_1$, $b_2$ and $b_3$ sectors
to the index ${\rm Tr}\,(-1)^{J'_0}$,
simply specify the number of chiral fermions contained in these sectors.
As an example of the second case, we now consider our second model with
$$
C{{b_1} \atopwithdelims[] {b_2}}=
C{{b_1} \atopwithdelims[] {b_3}}=-1, \quad
C{{b_2} \atopwithdelims[] {b_3}}=+1.
\numbereq\name{\eqmmb}
$$
We get $I_{b_1}=-16$ and $I_{b_2}=I_{b_3}=0$ from (\eqcb1) and (\eqccb23),
so we now have a model with only 16 generations instead. Again, this is
precisely what one would derive by analyzing the massless spectrum of this
model. In this model, the gauge group becomes
$SO(10)\times{SO(6)^3}\times SO(16)$, and none of the fermions that
would arise from the $b_2$ and $b_3$ sectors survive the GSO-projections,
so the $b_1$ sector is the only source of chiral fermions.

We now extend the basis ${\cal B}_1$ to
${\cal B}_2=\lbrace {\bf 1}, S, b_1, b_2, b_3, b_4, b_5 \rbrace$,
by adding two more basis vectors $b_4$
and $b_5$ of the following form [\fli1]
$$
\eqalign{
b_4&=(1\ 100\ 100\ 010\ 001\ 001\ 010 : 001001\ 000110\ 100\ 11111\
{\bf 0}_8),\cr
b_5&=(1\ 001\ 010\ 100\ 100\ 001\ 010 : 010001\ 100010\ 010\ 11111\
{\bf 0}_8)\cr} \numbereq\name{\eqb45}
$$
Since $N_{b_4}=N_{b_5}=2$,
the number of vectors in $\Xi^{-}$ for the models which are generated
by ${\cal B}_2$ increases to $M/2=2^7/2=64$.

In addition to the $16$ sectors in (\eqseta), the $48$ new sectors are:
$$
\eqalign{
&b_4\ (+\zeta),\ \ b_5\ (+\zeta),\ \ {\bf 1}+b_4+b_5\ (+\zeta),
\ \ S+b_4+b_5\ (+\zeta), \cr
&S+b_1+b_2+b_3+b_4\ (+\zeta),\ \ S+b_1+b_2+b_3+b_5\ (+\zeta), \cr
&b_1+b_4+b_5\ (+\zeta),
\ \ b_2+b_4+b_5\ (+\zeta),\ \ b_3+b_4+b_5\ (+\zeta), \cr
&S+b_1+b_2+b_4+b_5\ (+\zeta),\ S+b_2+b_3+b_4+b_5\ (+\zeta),\
S+b_3+b_1+b_4+b_5\ (+\zeta),\cr
&b_1+b_2+b_4(b_5)\ (+\zeta),\ \ b_2+b_3+b_4(b_5)\ (+\zeta),\ \
b_3+b_1+b_4(b_5)\ (+\zeta), \cr
&S+b_1+b_4(b_5)\ (+\zeta),\ \ S+b_2+b_4(b_5)\ (+\zeta),\ \
S+b_3+b_4(b_5)\ (+\zeta). \cr}
\numbereq\name{\eqsetb}
$$
Among these $48$ new sectors, only three sectors
$(b_4, b_5, S+b_4+b_5)$ provide massless space-time fermions.
It is straightforward to check that Eq. (\whys) holds
for any $\beta$ taken from the set (\eqsetb), so $S$ and $S+\zeta$
still do not contribute to the index (\eqboria). Similarly,
the contributions of $(b_1, b_2, b_3)$, all new
terms for $\beta$ from set (\eqsetb) vanish. However, because
$M/2$ increases by a factor of 4, (\eqcb1) and (\eqccb23) now
become
$$
I_{b_1}=2\biggl (\,C{{b_1} \atopwithdelims[] {b_2}}
+C{{b_1} \atopwithdelims[] {b_3}}\,\biggr ),\quad
I_{b_2}=8\biggl (\,C{{b_2} \atopwithdelims[] {b_3}}
+C{{b_2} \atopwithdelims[] {b_1}}\,\biggr ),\quad
I_{b_3}=8\biggl (\,C{{b_3} \atopwithdelims[] {b_1}}
+C{{b_3} \atopwithdelims[] {b_2}}\,\biggr ).
$$
Carrying out the same calculations for sectors $(b_4, b_5, S+b_4+b_5)$,
we have
$$
I_{b_4}=I_{b_5}=I_{S+b_4+b_5}=0.
\numbereq\name{\eqnn}
$$
Therefore, given the basis ${\cal B}_2$, the choice (\eqmma) gives
rise to $12$ generations, while the choice (\eqmmb) gives rise to
$4$ generations.

We would like to point out that the result (\eqnn) only depends on
the form of the basis vectors $b_4$ and $b_5$ and does not depend
on the choice of the spin-structure coefficients.
As a consequence no matter how one
chooses the relevant coefficients $C{{b_i} \atopwithdelims[] {b_j}}$,
in models generated by the basis ${\cal B}_2$, space-time
fermions arising from the sectors $(b_4, b_5, S+b_4+b_5)$ are {\it always}
vector-like. In fact, by analyzing the massless spectrum, one
can find that, the fermions from sectors $b_4$ or $b_5$ always
transform as ${\bf 16}+{\bf{\bar{16}}}$ of $SO(10)$, while those
from sector $S+b_4+b_5$ either transform as ${\bf 10}$ or as
singlets of $SO(10)$.

All the models we have considered above can be viewed upon
bosonization as symmetric $Z_2\times Z_2$ orbifold models, for a
discussion see {\it e.g.} Ref.~\ref{\moduli}{J. Lopez, D.V. Nanopoulos
and K. Yuan, \pr50 (1994) 4060.}. As we have mentioned in the beginning
of this section, these models in fact admit a $(2, 2)$ world-sheet
supersymmetry. To show that our index formula (\eqboria) in fact
works for some $(2, 0)$ models as well, we now would like to briefly
discuss one such example, the detailed analysis of this and other
models will be presented elsewhere \ref{\new}{I. Giannakis,
D. V. Nanopoulos and K. Yuan, CTP-TAMU preprint under preparation}.
In addition to the basis vectors in ${\cal B}_2$, we add one extra
basis vector of the form \ref{\search}{J. Lopez, D.V. Nanopoulos
and K. Yuan, \np399 (1993) 645.}.
$$
{\alpha}=(0\ 000\ 000\ 000\ 000\ 000\ 011 : 000001\ 011001\
\h\h\h\ \h\h\h\h\h\ \h\h\h\h\ 1100).
\numbereq\name{\eqrigitu}
$$
This new basis vector breaks down the right-moving $(0, 2)$ structure,
and has order $N_\alpha=4$. With the basis
${\cal B}_3=\lbrace {\bf 1}, S, b_1, b_2, b_3, b_4, b_5, \alpha \rbrace$,
the set $\Xi^{-}$ contains $M/2=256$ vectors, namely, the 64 vectors
of (\eqseta) and (\eqsetb) and those which can be obtained by
adding to them $\alpha$, $2\alpha$ and $3\alpha$. Since this
basis contains a $(2, 2)$ ``core'', we can still use the 12 real
fermions ${\bar y}^I, {\bar \omega}^I$ and the three complex fermions
${\bar \eta}^{1,2,3}$ as the right-moving internal degrees of
freedom in our expression for the index
(\eqboria), and carry out the calculation
essentially the same way. However, because of the absence of the
$(0, 2)$ structure in the current case, the analysis turns out to
be somewhat involved. For one thing, the contribution of each
sector to the $N=2$ index, namely, each term in the $\alpha$-summation
of (\eqboria) by itself is no-longer bound to be an integer. In spite
of this fact, the cancellations which take place in the $\alpha$-summation
always render the index to be an integer. Futhermore, the role of the
spin-structure coefficients $C{{b_i} \atopwithdelims[] {b_j}}$ becomes
more important. For the choice given in Ref.~[\search], we get [\new]
$$
{\rm Tr}\,(-1)^{J'_0}=-3
\numbereq\name{\eqkcos}
$$
This is in complete agreement with the result that this model has
three generations [\search].

\newsection Conclusions.

In this last section we will summarize the results presented in this
paper. The free fermionic string models represent a class of classical
string vacua. Although they lack a geometric interpretation, they are
particularly simple to build and as a result several realistic models
have been constructed.

In string theory there seems to be an intimate connection
between spacetime and
world-sheet physics. In the spirit of this philosophy
and in the context
of free fermionic string models, we have associated the
number of generations of a string vacuum with $N=1$
spacetime supersymmetry
with the index of the supersymmetry generator of the
underlying $N=2$ internal superconformal theory. The existence of
$N=1$ spacetime supersymmetry appears to be vital in our attempts to
define the index in terms of the zero mode of the
$U(1)$ Kac-Moody current of the $N=2$ algebra
and to interpret it as the number of
generations of a string vacuum.
The index depends on the internal fermions and the
formula we derived
in section $3$ is sensitive only
to the boundary conditions of the internal fermions and
the coefficients
which represent the weights with which each sector contributes to
the one-loop partition function. We have
also applied this formula to a number of realistic free fermionic models
and have calculated the number of generations $N_g$ of these models.
Furthermore we have verified our results by explicitly deriving the
massless spectrum of these vacua.
The index being a topological quantity is invariant
under deformations of the superconformal
structure. Thus, it characterizes
a family of string vacua rather than a particular vacuum
of string theory.

Finally one last observation is in order. The
different sectors of the models
contribute to the formula for the index powers of $2$. The
origin of this phenomenon resides in the
equivalence of free fermionic models with the $Z_2 \times Z_2$ orbifold
compactifications. The number of
generations in orbifold models is related to
the Euler character of the orbifold which equals
the number of its fixed points (powers of 2 for $Z_2$ orbifolds).
A generic orbifold model typically produces a large
number of generations. In order to reduce
the number one incorporates
either the action of a discrete group with no fixed points or introduces
Wilson lines.
On the contrary in free fermionic models the number of generations
is sensitive only to the boundary conditions of the fermions and
to certain coefficients.
By introducing extra basis vectors compatible with the necessary
rules we can reduce the number of generations. As a result
it is easier to incorporate their effect into a general formula
like the one we have presented in section $3$.

\newsection Acknowledgments.

We would like to thank I. Antoniadis and M. Evans for useful discussions.
This work has been supported in part by DOE grant DE-FG05-91-ER-40633.
\nobreak\bigskip

\immediate\closeout1
\bigbreak\bigskip

\line{\bf References.\hfil}
\nobreak\medskip\vskip\parskip

\input refs

\vfill\end

\bye